\magnification=1200

\let\a=\alpha \let\f=\varphi 
\let\d=\delta   \def\d{\delta} \let\s=\sigma
   \let\Si=\Sigma

\let\*=\medskip \def\\{\hfill\break} \def\fra#1#2{{#1\over#2}}
\def\XX{{\cal X}}  
\def\grad{\,{\rm grad}\,}\def\div{\,{\rm div}\,} 

\null\vskip 4truecm
\centerline{SRB STATES AND NONEQUILIBRIUM STATISTICAL}
\centerline{MECHANICS CLOSE TO EQUILIBRIUM.} 
\*\*\*\*
\centerline{by Giovanni Gallavotti\footnote{*}{Permanent address:
Fisica, Universit\'a di Roma ``La Sapienza'', 00185 Roma, Italia} and
David Ruelle\footnote{**}{Permanent address: IHES, 91440 Bures sur
Yvette, France.}}
\medskip
{\bf Abstract.} {\it Nonequilibrium statistical mechanics close to
equilibrium is studied using SRB states and a formula [10] for their 
derivatives with respect to parameters.  We write general expressions
for the thermodynamic fluxes (or currents) and the transport
coefficients, generalizing the results of [4], [5].  In this framework
we give a general proof of the Onsager reciprocity relations.}
\medskip
	There is currently a strong revival of nonequilibrium
statistical mechanics, based on taking seriously the nonlinear (chaotic)
microscopic dynamics.  One natural idea in this direction is to
use nonequilibrium steady states, which are defined to be the SRB
states for the dynamics (see below).

	The idea of using SRB states eventually led to useful results only 
recently, when it was combined in [6] with 
reversibility of the dynamics to obtain a nontrivial fluctuation
formula for the entropy production. In [5] the same ideas were applied
to prove the Onsager reciprocity relations, and the
fluctuation-dissipation formula for a rather special
class of models.  The analysis dealt with examples rather than the
general situation, and relied on an
unproven conjecture on Anosov systems.  In this note
we generalize [5] and, using [10], give results that can be proved 
rigorously for {\it Axiom A} diffeomorphisms (this is a strong 
chaoticity assumption, see [12], here we skip technical details).

	To be discussed is the classical microscopic description of the
time evolution of a physical system.  (No {\it large system} assumption
will be made).  We let the accessible phase space be a compact manifold
$M$.  The time evolution is given by iterates of a diffeomorphism $f$ of
$M$ (discrete time case) or by integrating a vector field $F$ on $M$
(continuous time case).  We choose a Riemann metric on $M$,
and let $dx$ be the corresponding volume element.

	In nonequilibrium statistical mechanics, the time evolution
typically does not preserve any measure which has a density with respect to
$dx$.  Let $m(dx)=\underline m(x)dx$ be a probability measure (with
density $\underline m$), and $f^{*k}m$ the direct image of $m$ by
$f^{*k}$; any weak limit for $n\to\infty$ of

$$ {1\over n}\sum_{k=0}^{n-1}f^{*k}m $$ 
is an $f$-invariant probability measure $\rho_f$ on $M$.  If furthermore
$\rho_f$ is ergodic we may say that it is a {\it natural nonequilibrium state}
(the SRB states are special examples of this, [3]).  

         An infinitesimal change $\delta f$ of $f$ corresponds to an 
(infinitesimal) vector field $X=\delta f\circ f^{-1}$ on $M$ and an easy
formal calculation gives

$$	\delta\rho_f(\Phi)=\sum_{n=0}^\infty\rho_f
	\langle{\rm grad}(\Phi\circ f^n),X\rangle\eqno(1)     $$
where $\Phi$ is a smooth test function and
$\langle\cdotp,\cdotp\rangle$ is the scalar product of a tangent vector
and a cotangent vector to $M$.  Equation (1) expresses the
change in the expectation value of the observable $\Phi$ when the system
is subjected to a force $X$.  Replacing $X$ by a time dependent force
$X_t$ we have for the change of expectation value of the observable
$\Phi$ at time $s$ the analogous formula

$$	\delta_s\rho_f(\Phi)=\sum_{t\le s}\rho_f
	\langle{\rm grad}(\Phi\circ f^{-t}),X_t\rangle\eqno(2)     $$

Note that the condition $t\le s$ corresponds to the "causality
principle" and (2) can be used to derive Kramers-Kronig dispersion
relations.  For Axiom A diffeomorphisms, (1) and (2) can be proved
rigorously\footnote{*}{See [10]; the proof given in [10] for the time
dependent case assumes that the pertubation has finite support in
time.}: the left-hand side is a derivative, and the right-hand side a
convergent series.

	In the case of continuous time systems described by a
differential equation $\dot x= F(x)$ and by the corresponding flow $(f^t)$,
an infinitesimal variation $\delta F=X$ generates a
variation in the expectation value $\rho_F(\Phi)$ (for the {\it
natural nonequilibrium state}) given by:

$$ \delta\rho_F(\Phi)=\int_0^{+\infty} dt\int \rho_F(dx)
\langle\grad_x(\Phi\circ f^t),X(x)\rangle\eqno(1') $$
(A rigorous proof for Axiom A flows has not been given yet).
\medskip
	The entropy production associated with the diffeomorphism $f$
is defined by\footnote{**}{For a discussion
of entropy production see [9].}

$$	e_f=\rho_f(\sigma_f)\qquad,\qquad\sigma_f=-\log J_f\eqno(3)     $$
where $J_f$ is the absolute value of the Jacobian of $f$ with respect to
the Riemann volume element $dx$.  In the continuous time systems, we let

$$ 	e_F=\rho_F(\sigma_F)\qquad,\qquad\sigma_F=-{\rm div}F\eqno(3')$$
(see [5]).

	From now on we shall fix $f$ such that $\rho_f(dx)=dx$, {\it i.e.},
$\rho_f$ is the Riemann volume element (if $\rho_f$ has smooth density
this can be achieved by a change of metric).  Note that in particular
this implies that (1) can be rewritten as

$$	\delta\rho_f(\Phi)=\sum_{n=0}^\infty\rho_f
	(\Phi\circ f^n\ldotp(-{\rm div}X))\eqno(4)     $$

In this {\it nondissipative} situation we have $\log J_f=0$, hence
$e_f=0$.  If we write as before $X=\delta f\circ f^{-1}$ we obtain to
second order in $X$, using (3) and (1),
$$	e_{f+\delta f}=(\rho_f+\delta\rho_f)(\sigma_{f+\delta f})     $$
$$	={1\over 2}\rho_f(({\rm div}X)^2)-\delta\rho_f({\rm div}X\circ f)     $$
$$	={1\over 2}\rho_f(({\rm div}X)^2)
-\sum_{n=0}^\infty\rho_f\langle{\rm grad}(({\rm div}X)\circ f^n),X\rangle $$
$$	={1\over 2}\rho_f(({\rm div}X)^2)
+\sum_{n=1}^\infty\rho_f(({\rm div}X)\circ f^n\ldotp{\rm div}X)\eqno(5)     $$
$$	={1\over 2}\sum_{n=-\infty}^\infty\rho_f(({\rm div}X)\circ f^n
	\ldotp{\rm div}X)     $$

	The same analysis leads, in the continuous time case (with 
$\delta F=X$, and $\rho_F(dx)=dx$) to:

$$e_{F+\d F}=\fra12 \int_{-\infty}^{+\infty}dt\,\rho_f(\s_f
\circ f^t\cdot\s_f)   $$

$$  =\fra12\int_{-\infty}^{+\infty} dt\int dx\,\div X(f^tx)\,\cdot\,\div
X(x)\eqno(5')$$
see [4], [5].
\medskip
	We shall now relate the above expressions for the entropy
production to the definition
of the {\it thermodynamic forces} $\XX_\a$, and the {\it conjugated
thermodynamic fluxes} ${\cal J}_\alpha$ as they appear for instance in
[8].  We begin by an informal discussion, and assume, as usual in
applications, that $f$ or $F$ depends on parameters $E_\alpha$, so
that we may write (to first order)
$$	X=\sum_\alpha V_\alpha\delta E_\alpha     $$
We identify the thermodynamic forces ${\cal X}_\alpha$ with
the parameters $E_\alpha$.  Considering first the continuous time
case, we follow [4], [5], and we define the thermodynamic flux
conjugated to $E_\alpha$ as 
$$	{\cal J}_\alpha
	=\rho_{F+\delta F}({\partial\over\partial E_\alpha}
	\sigma_{F+\delta F})     $$
Since $\partial\sigma/\partial E_\alpha$ is a divergence, 
$\rho(\partial\sigma/\partial E_\alpha)=0$ and we have
$$	{\cal J}_\alpha
	=\delta\rho_F(-{\rm div}V_\alpha)\qquad+{\rm h.o.}   $$
$$	=\int_0^{+\infty}dt\int\rho_F(dx) 
\langle{\rm grad}_x((-{\rm div}V_\alpha)\circ f^t),X(x)\rangle
	\qquad+{\rm h.o.}   $$
 From now on we neglect higher order terms and (using integration by
parts, since $\rho_F(dx)=dx$) we write
$$	{\cal J}_\alpha
=\int_0^{+\infty}dt\int\rho_F(dx)\,({\rm div}_xX)({\rm div}_{f^tx}V_\alpha)
$$

	In the discrete time case we define the thermodynamic flux only to
leading order in the $E_\alpha$ by 
$$	{\cal J}_\alpha
	={1\over2}\rho_f(({\rm div}V_\alpha)({\rm div}X))
+\sum_{n=1}^\infty\rho_f(({\rm div}V_\alpha)\circ f^n\ldotp({\rm
div}X))$$

	With these definitions ${\cal J}_\alpha$ depends only on the
application of $X$ in the past (causality, cf (2)) and the entropy
production (to second order) is
$$	e_{f+\delta f}\qquad{\rm or}\qquad e_{F+\delta F}
	=\sum_\alpha{\cal X}_\alpha{\cal J}_\alpha     $$
These conditions uniquely determine the ${\cal J}_\alpha$.  Notice
that the formulae for ${\cal J}_\alpha$ involve only the divergences
of $X$ and $V_\alpha$.
\medskip
	To continue the discussion, we assume that there is a
(sufficiently large) Banach space ${\cal B}$ of functions
$\Phi:M\to{\bf R}$ such that

$$   \rho_f(\Phi)=0\qquad\qquad{\rm if}\qquad\qquad\Phi\in{\cal B}   $$
and for some constant $C$

$$	\sum_{k\in{\bf Z}}|\rho_f(\Psi\circ f^k\ldotp\Phi)|\
	\le C\|\Phi\|_{\cal B}\|\Psi\|_{\cal B}
	\qquad\qquad{\rm if}\qquad\qquad\Phi,\Psi\in{\cal B}\eqno(6)     $$
[This is the discrete time case, the continuous time case is similar.
If $f$ is an Anosov diffeomorphism\footnote{*}{An Axiom A
diffeomorphism $f$ preserving $dx$ is an Anosov diffeomorphism.} we
can take for ${\cal B}$ a space of H\"older continuous functions on
$M$.  Similarly for Anosov flows].	

	From now on, we assume that ${\rm div}X$ is in the Banach space ${\cal
B}$ just introduced, and we may define ${\cal X}\in{\cal B}$ and ${\cal
J}\in{\cal B}^*$ (the dual of ${\cal B}$) as follows:
$$	{\cal X}=-{\rm div}X\in{\cal B}\eqno(7)     $$
$$	({\cal J},\Phi)={1\over 2}\rho_f(-{\rm div}X\ldotp\Phi)
	+\sum_{n=1}^\infty\rho_f((-{\rm div}X)\circ
f^n\ldotp\Phi)\eqno(8)$$
for discrete time, and
$$	({\cal J},\Phi)={1\over 2}\int_0^\infty\rho_F((-{\rm div}X)\circ
f^t\ldotp\Phi)\eqno(8')$$
for continuous time, where $\Phi\in{\cal B}$, and 
$(\cdotp,\cdotp)$ is the pairing ${\cal B}^*\times{\cal B}\to{\bf C}$.
Note that if ${\rm div}V_\alpha\in {\cal B}$ we have
$$	 {\cal J}_\alpha=({\cal J},-{\rm div}V_\alpha)     $$
and that the entropy production is
$$	e_{f+\delta f}\qquad{\rm or}\qquad e_{F+\delta F}
	=({\cal J},{\cal X})     $$
\medskip
	With the above notation and assumptions we may write 
${\cal J}=L{\cal X}$,
where $L$ is, in view of (6), a continuous linear map ${\cal B}\to{\cal
B}^*$.  If we define a unitary operator $U$ on $L^2(\rho_f)$ by
$U\Phi=\Phi\circ f$ and write $U=\int_{-\pi}^\pi e^{i\alpha}\, d{\bf
P}(\alpha)$ we have
$$	\sum_{k\in{\bf Z}}e^{-ik\alpha}\rho_f(\Phi\circ f^k\ldotp\Phi)
	=2\pi{d\over{d\alpha}}(\Phi,d{\bf P}(\alpha)\Phi)_{L^2}     $$
so that  the quadratic form associated with $L$ satisfies
$$	(L\Phi,\Phi)={1\over 2}\sum_{k\in{\bf Z}}\rho_f(\Phi\circ f^k\ldotp\Phi)
=\pi{d\over{d\alpha}}(\Phi,d{\bf P}(\alpha)\Phi)_{L^2}\big|_{\alpha=0}\ge0$$
In particular, this quadratic form is $\ge0$.
\medskip
	The formulae obtained up to now hold under the only assumption
of closeness to equilibrium.  If we make the further assumption of
(microscopic) reversibility, we shall obtain a symmetry property of
$L$ called Onsager reciprocity.  For simplicity we discuss only the
discrete time case.

	We say that the dynamics is {\it reversible} if there exists a
diffeomorphism $i:M\to M$ such that $i^2={\rm identity}$, $i\circ
f=f^{-1}\circ i$.  We have then also $i^*\rho_f=\rho_f$.  [Note that
$\rho_f$, {\it i.e.}, the Riemann volume, is mixing by (6), hence
$f$-ergodic.  Since $i^*\rho_f$ is absolutely continuous with respect
to $\rho_f$ and satisfies $f^*(i^*\rho_f)=i^*(f^{-1})^*\rho_f=(i^*\rho_f)$
we have $i^*\rho_f=\rho_f$ by ergodicity].  Assuming reversibility
we may define $\epsilon\Phi=\Phi\circ i$ for $\Phi\in{\cal B}$, and we find
$$	(L\Psi,\Phi)={1\over 2}
\rho_f(\Psi\ldotp\Phi)+\sum_{n=1}^\infty\rho_f(\Psi\circ f^n\ldotp\Phi) 
	={1\over 2}\rho_f(\epsilon\Psi\ldotp\epsilon\Phi)
+\sum_{n=1}^\infty\rho_f(\epsilon\Psi\circ f^{-n}\ldotp\epsilon\Phi)  $$
$$	={1\over 2}\rho_f(\epsilon\Phi\ldotp\epsilon\Psi)
+\sum_{n=1}^\infty\rho_f(\epsilon\Phi\circ f^{-n}\ldotp\epsilon\Psi)
	=(L(\epsilon\Phi),\epsilon\Psi)     $$
The relation
$$	(L\Psi,\Phi)=(L(\epsilon\Phi),\epsilon\Psi)     $$
is a form of the {\it Onsager reciprocity relation} as we shall see
in a moment.  Note that reversibility was assumed only for $f$ ({\it
i.e.}, at equilibrium), the perturbation $\delta f$ is arbitrary.
\medskip
	To obtain a more familiar form of the entropy production formula (see
[8]), we assume that ${\cal B}$ has a basis $(\Phi_\alpha)$ with a
corresponding system $(\phi_\alpha)$ of elements of ${\cal B}^*$ such
that $(\phi_\alpha,\Phi_\beta)=\delta_{\alpha\beta}$ (see [11]).  Write
$$	{\cal X}_\alpha=(\phi_\alpha,{\cal X})=(\phi_\alpha,-{\rm div}X)     $$
$$	{\cal J}_\alpha=({\cal J},\Phi_\alpha)
={1\over2}\rho_f(-{\rm div}X,\Phi_\alpha)
+\sum_{n=0}^\infty\rho_f((-{\rm div}X)\circ f^n\ldotp\Phi_\alpha)     $$
In particular
$$	-{\rm div}X=\sum_\alpha(\phi_\alpha,-{\rm div}X)\ldotp\Phi_\alpha
	=\sum_\alpha{\cal X}_\alpha\ldotp\Phi_\alpha     $$
and
$$	e_{f+\delta f}=({\cal J},{\cal X})
	=\sum_\alpha{\cal J}_\alpha\ldotp{\cal X}_\alpha     $$
$$	{\cal J}_\alpha
	={1\over 2}\sum_\beta{\cal X}_\beta\rho_f(\Phi_\beta\Phi_\alpha)
	+\sum_{n=1}^\infty\sum_\beta{\cal X}_\beta
	\rho_f(\Phi_\beta\circ f^n\ldotp\Phi_\alpha)     $$
To avoid convergence problems suppose that finitely many ${\cal
X}_\beta$ only are nonzero.  Then
$$	{\cal J}_\alpha=\sum_\beta L_{\alpha\beta}{\cal X}_\beta     $$
where
$$	L_{\alpha\beta}={1\over 2}\rho_f(\Phi_\beta\Phi_\alpha)
	+\sum_{n=1}^\infty\rho_f(\Phi_\beta\circ f^n\ldotp\Phi_\alpha)  $$
\medskip
	Suppose again reversibility of the dynamics, and let the basis
$(\Phi_\alpha)$ of ${\cal B}$ be such that
$\epsilon\Phi_\alpha=\Phi_\alpha\circ f=\epsilon_\alpha\Phi_\alpha$ with
$\epsilon_\alpha=\pm 1$.  Then
$$	L_{\alpha\beta}=\epsilon_\alpha\epsilon_\beta L_{\beta\alpha}	$$
which is the usual form of the Onsager reciprocity relation.
\medskip
	We conclude by sketching an example, see [1], of the
formalism described above.  Let $\Sigma$ be a surface of constant
negative curvature, and genus $g$, with the $g-1$ automorphic forms
$\phi_\alpha(z)dz$.  We normalize these forms so that they are
orthonormal in the space $L_2(T\Si)$ in the natural scalar product, [7]. 

We can then consider the hamiltonian equations of the motion of a
particle on $\Si$ subject to the external force generated by the
``electric'' field ${\cal E}$ such that ${\cal E}_xdx+{\cal
E}_ydy={\rm Re}\,\sum E_\a\f_\a(z)dz$.

We also impose, via Gauss' principle, [5], that there is a {\it thermostat}
force that keeps the kinetic energy constant (and equal to $1/2$) in spite of 
the field's action. Thus the ``only'' efffect that the fields will have on
the flow is that currents flowing ``around'' the $g$ ``holes'' of the
surface will be established. But the flow being a mixing Anosov flow on a 
compact surface (because of the gaussian constraint) it will result that
a stationary state will be reached and the latter will be the SRB
distribution, [2].

The equations of motion can be written explicitly and one finds in
particular that the entropy creation rate at the point
$\vec q,\vec p$ of the phase space is

$$\s(\vec q,\vec p)=\fra{\vec{\cal E}\cdot\vec p}{p^2}$$
The transport coefficient can also be explicitly computed, and one
finds $L_{\alpha\beta}={1\over 2}\delta_{\alpha\beta}$, see [1].
\*\*\*
	{\bf References.}
\*

[1] F.Bonetto, G.Gentile, V.Mastropietro. ``Electric
fields on a surface of constant negative curvature.'' Preprint, 1996.

[2] R.Bowen, D.Ruelle. ``Ergodic theory of Axiom A flows.'' 
Inventiones Math. {\bf 29},181-202(1975).

[3] J.-P.Eckmann and D.Ruelle. ``Ergodic theory of chaos and strange
attractors'' Rev. Mod. Phys. {\bf 57},617-656(1985).

[4] G.Gallavotti: ``Extension of Onsager's reciprocity to large fields and
the chaotic hypothesis.'' Phys. Rev. Letters. {\bf
77},4334-4337(1996).

[5] G.Gallavotti.  "Chaotic hypothesis: Onsager reciprocity and
fluctuation-dissi\-pa\-tion theorem."  J. Statist. Phys. 
{\bf 84}, 899--926(1996)

[6] G.Gallavotti and E.G.D.Cohen.  "Dynamical ensembles in nonequilibrium
statistical mechanics."  Phys. Rev. Letters {\bf 74},2694-2697(1995).
"Dynamical ensembles in stationary states."  J. Statist. Phys.
{\bf 80},931-970(1995).

[7] I.M.Gel'fand, M.I.Graev. I.I.Pyateckii-Shapiro.  {\it
Representation theory and automorphic functions.}  Saunders,
Philadelphia, 1969.

[8] S.R.de Groot and P.Mazur.  {\it Non-equilibrium statistical
thermodynamics.}  Do\-ver, New York, 1984.

[9] D.Ruelle.  "Positivity of entropy production in nonequilibrium
statistical mechanics."  J. Statist. Physics, {\bf 85},1-25(1996).

[10] D.Ruelle.  "Differentiation of SRB states."  To be published; the
preprint is archived in {\tt mp$\_$arc@math.utexas.edu}, {\bf96}, \#499.

[11] I.Singer.  {\it Bases in Banach spaces I.}  Springer, Berlin, 1970.

[12] S.Smale.  "Differentiable dynamical systems."  Bull. Amer. Math. Soc.
{\bf 73},747-817(1967).	

\end